\documentstyle[aps,prl,epsf,multicol]{revtex}
\makeatletter
\newlength\abovecaptionskip \newlength\belowcaptionskip
\def\@makecaption#1#2{%
 \vskip\abovecaptionskip \sbox\@tempboxa{#1: #2}%
 \ifdim \wd\@tempboxa >\hsize #1: #2\par \else \global \@minipagefalse
 \hb@xt@\hsize{\hfil\box\@tempboxa\hfil}%
 \fi \vskip\belowcaptionskip} \makeatother

\begin{document}
\title{Extended Fractons and Localized Phonons on Percolation Clusters}
\author{Jan W. Kantelhardt$^1$, Armin Bunde$^1$, and 
 Ludwig Schweitzer$^2$}
\address{$^1$ Institut f\"ur Theoretische Physik, 
 Justus-Liebig-Universit\"at Giessen, Heinrich-Buff-Ring 16, 
 D-35392 Giessen, Germany}
\address{$^2$ Physikalisch-Technische Bundesanstalt, 
Bundesallee 100, D-38116 Braunschweig, Germany}
\date{Received 20. July 1998}
\draft
\maketitle
\begin{multicols}{2}[%
\begin{abstract}
We investigate the localization behavior of vibrational modes of infinite
percolation clusters above the critical concentration in two and three
dimensions using the method of level statistics.  We find that all
eigenstates are localized in $d=2$, including the low frequency phonon
states.  In $d=3$ we obtain evidence for a localization-delocalization
transition.  But contrary to the common view this transition occurs for
frequencies above the phonon fracton crossover giving rise to a new regime
of extended fracton states.
\end{abstract}
]

Percolation represents one of the standard models for disordered systems.
Its applications range from transport in composites and in amorphous and
porous media to the properties of branched polymers and gels, for recent
reviews see e.g.  \cite{stauffer,bunde96}.  One intriguing aspect is the
vibrational dynamics of percolation networks.  Since the pioneering
discovery of localized lattice vibrations (``fractons'') by Alexander and
Orbach \cite{alexander,orbach}, much work has been devoted to the study of
their properties \cite{havlin,grest,southern,yakubo,devries,lambert,%
bunde92,petri,argyrakis,kantelhardt97} and the investigation of various
applications \cite{tsujimi,montagna,stoll,russ}; see \cite{nakayama} for 
a recent review.

It is well accepted, that all vibrational excitations are localized at the
percolation threshold, and the density of states (DOS) scales as $g(\omega)
\sim \omega^{d_{\rm s}-1}$, where $d_{\rm s} \approx 4/3$ is the spectral
dimension \cite{alexander,orbach,havlin,grassberger}.  Above criticality,
there exists a crossover frequency $\omega_\xi$:  for frequencies larger
than $\omega_\xi$, $g(\omega) \sim \omega^{d_{\rm s}-1}$, while below
$\omega_\xi$, the DOS scales in the normal fashion, $g(\omega) \sim
\omega^{d-1}$.  The meaning of the ``fracton-phonon crossover'' at
$\omega_\xi$ and in particular its implication on the localization behavior
has been discussed in a controversial way.  Analogies with random walks
seem to indicate that the crossover characterizes a transition between
localized fractons and extended phonons.  This view has been mostly
accepted for three-dimensional systems (\cite{stauffer,bunde96,orbach,%
grest,argyrakis,montagna,nakayama}, but see also \cite{tsujimi}), but not
yet been proven.  Analogies with the Anderson model \cite{rammal}, in
contrast, suggest the absence of a localization-delocalization transition
for two-dimensional systems.  But since these analogies are not exact, even
the value of the critical dimension has not been established for vibrations
\cite{petri}.

It is the purpose of this Letter to clarify the localization properties of
the vibrational excitations.  We employ the method of level statistics
which has been used successfully to determine the
localization-delocalization transition (LDT) in the Anderson model
\cite{SEA93,HS94a,BSZK96}.  We find that the localization properties of the
vibrations are not related to the fracton-phonon crossover (FPC) at
$\omega_\xi$.  Our results suggest that ({\it i\/}) in $d=2$ all states are
localized and ({\it ii\/}) the LDT in $d=3$ occurs at a critical frequency
$\omega_{\rm c}$ well above $\omega_\xi$.  Accordingly the vibrational
modes in $d=3$ are extended phonons below $\omega_\xi$ and localized
fractons above $\omega_{\rm c}$, while the regime between $\omega_\xi$ and
$\omega_{\rm c}$ is characterized by extended excitations with a
fracton-type dispersion (``extended fractons'').

We consider the incipient infinite site percolation cluster on a square
lattice and a simple cubic lattice, both above the critical concentrations
$p_{\rm c} \simeq 0.59275$ and $\simeq 0.3116$, respectively.  On length
scales below the correlation length $\xi \sim \vert p-p_{\rm
c}\vert^{-\nu}$, the cluster is self similar and characterized by the
fractal dimension $d_f$.  Above $\xi$, the cluster appears homogeneous.
The correlation exponent $\nu$, the fractal dimension $d_f$, and the
spectral dimension $d_{\rm s}$ depend on the space dimension $d$:  $\nu =
4/3$, $d_f = 91/48$ and $d_{\rm s} \simeq 1.3173$ in $d=2$
\cite{grassberger}; $\nu \simeq 0.875$, $d_f \simeq 2.524$, and $d_{\rm s}
\simeq 1.328$ in $d=3$ \cite{stauffer,bunde96}.

We assume that equal masses $M$ are placed on each occupied site and
nearest neighbor cluster sites $n$ and $n+\delta$ are coupled by equal
(scalar) force constants $k_{n,n+\delta}$.  In this case, different
components of displacements decouple and we obtain the same equation of
motion for lattice vibrations for all components $u_n(t)$,
\begin{equation}
 M \; {d^2 \over dt^2} u_n(t) = {\sum_\delta}^{\prime}\,
 k_{n,n+\delta} \left[ u_{n+\delta}(t) - u_n(t) \right],
\label{eq:frakza} \end{equation}
where the sum runs over all nearest neighbor sites $n+\delta$ of site $n$.
The standard ansatz $u_n(t) = u_{n, \omega} \; \exp(-i\omega t)$ leads to
the corresponding time independent vibration equation.  Setting
$M=k_{n,n+\delta}=1$ (which determines the unit of the frequency) we obtain
\begin{equation}
 -\omega^2 \; u_{n,\omega} = {\sum_\delta}^{\prime}\,
 (u_{n+\delta,\omega} - u_{n,\omega}),
\label{eq:frak} \end{equation}
which represents an eigenvalue equation that can only be solved
numerically.

For obtaining an analytical estimate of the way the characteristic length
in the vibration problem depends on the frequency $\omega$ and on the
correlation length $\xi$, one can employ the analogy
\cite{alexander,orbach,havlin} between the corresponding diffusion equation
and the vibration equation, Eq.~(\ref{eq:frakza}).  In the diffusion
problem, the characteristic length is the root-mean-square displacement
$R(t)=\langle r^2(t) \rangle^{1/2}$ of a diffusing particle (random
walker).  At criticality, $R(t)$ scales as $R(t) \sim t^{1/d_{\rm w}}$ with
the random walk dimension $d_{\rm w} = 2d_f /d_{\rm s}$.  Above $p_c$, the
finite correlation length $\xi$ gives rise to a finite `correlation' time
$t_\xi \sim \xi^{d_{\rm w}}$.  Below $t_\xi$, the random walker discovers
the fractal part of the infinite cluster and $R(t) \sim t^{1/d_{\rm w}}$,
while above $t_\xi$, on length scales where the cluster is homogeneous,
diffusion is normal and $R(t) \sim t^{1/2}$.  Since the (discretized)
diffusion equation reduces to the vibration equation,
Eq.~(\ref{eq:frakza}), when the first time derivative is substituted by a
second time derivative, we can obtain a corresponding characteristic length
$\Lambda(\omega)$ for the vibration problem by substituting $t^{-1}$ in
$R(t)$ by $(1/t)^2 = \omega^2$.  This yields \cite{alexander,grest}
\begin{equation}
  \Lambda(\omega) \sim \cases{
    \omega^{-2/d_{\rm w}}, & for $\omega \gg \omega_\xi \sim 
	t_\xi^{-1/2} \sim  \xi^{-d_{\rm w}/2}$, \cr 
    \omega^{-1}, & for $\omega \ll \omega_\xi$.}
\label{eq:lambda} \end{equation}
The expression for the crossover frequency $\omega_\xi$ that separates the
fracton from the phonon regime can be written as
\begin{equation}
 \omega_\xi = \omega_0 \; (p-p_{\rm c})^{d_{\rm w}\nu/2},
\label{eq:omegaxi} \end{equation}
where the proportionality factor $\omega_0$ depends on the details of the
underlying lattice.

The scaling for the vibrational DOS $g(\omega)$ can be derived in an
analogous way.  By construction, the crossover frequency $\omega_\xi$
separating the fracton regime with $g(\omega) \sim \omega^{d_{\rm s}-1}$
from the phonon regime with $g(\omega) \sim \omega^{d-1}$ is expected to be
identical with $\omega_\xi$ from Eq.~(\ref{eq:omegaxi}).  We have
determined $\omega_0$ from the DOS for percolation clusters on the square
lattice and the simple cubic lattice for several concentrations above
$p_{\rm c}$.  The crossover from the fracton to the phonon regime is 
smooth (see also \cite{nakayama}).  We find $\omega_0 \approx 15$
in $d=2$ and $\omega_0 \approx 6$ in $d=3$ for the parameter in
Eq.~(\ref{eq:omegaxi}).  Both numbers are in good agreement with earlier
(less extensive) numerical calculations of the DOS \cite{nakayama}.

\begin{figure}\centering
\epsfxsize8.4cm\epsfbox{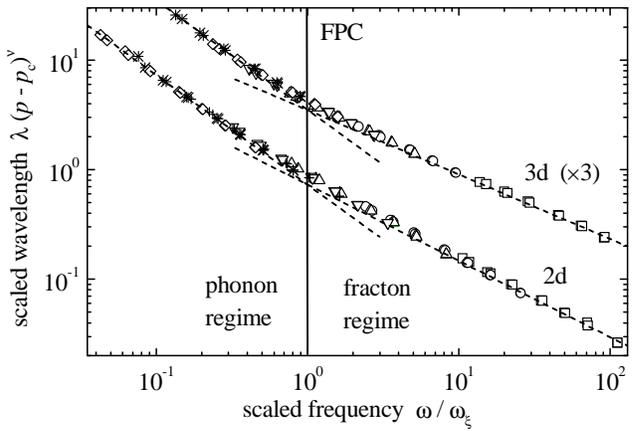}
\parbox{8.5cm}{\caption[]{\small 
Scaled wavelength $\lambda(\omega) (p-p_{\rm c})^\nu$ versus scaled
frequency $\omega /\omega_\xi$ for site percolation clusters in $d=2$ and
$d=3$ with $\omega_\xi$ from Eq.~(\protect{\ref{eq:omegaxi}}) and
$\omega_0$ determined from the scaling behavior of the DOS.  The upper
curve ($d=3$) has been shifted by a factor of three.  The symbols
correspond to the concentrations of occupied sites, for $d=2$:  $p=0.61$
($\Box$), $0.63$ ({\Large$\circ$}), $0.66$ ($\bigtriangleup$), $0.70$
($\bigtriangledown$), $0.75$ ($+$), $0.82$ ($\star$), and $0.90$
($\Diamond$); for $d=3$:  $p=0.32$ ($\Box$), $0.35$ ({\Large$\circ$}),
$0.37$ ($\bigtriangleup$), $0.40$ ($\bigtriangledown$), $0.44$
($\Diamond$), and $0.50$ ($\star$).  The straight lines are fits to the
data with slope $-1$ in the phonon regime ($\omega < \omega_\xi$) and
$-2/d_{\rm w}$ in the fracton regime ($\omega > \omega_\xi$) according to
Eq.~(\protect{\ref{eq:lambda}}).  In $d=3$, we used the effective value
$d_{\rm w}=3.4$ that governs diffusion on length scales of the order of
the correlation length considered here.  The asymptotic value
$d_{\rm w}=3.8$ applies only to considerably larger length scales.
The figure shows that $\omega_\xi$ is the FPC frequency for the
wavelength.}
\label{fig:wl}}
\end{figure}

In order to see the significance of the characteristic length $\Lambda$
described by Eq.~(\ref{eq:lambda}), we have numerically solved
Eq.~(\ref{eq:frak}) using the Lanczos algorithm.  Figure \ref{fig:wl} shows
the wavelength $\lambda(\omega)$ that we calculated by Fourier transform of
the eigenstates at several frequencies and concentrations in $d=2$ and in
$d=3$.  The data nicely confirm that the wavelength $\lambda$ is the
characteristic length $\Lambda$ discussed above.  At the FPC, the
wavelength crosses over from fracton behavior for high frequencies to
phonon behavior for low frequencies according to Eq.~(\ref{eq:lambda}).
The values of the crossover frequency $\omega_\xi$ are well described by
Eq.~(\ref{eq:omegaxi}).  In addition, the prefactor $\omega_0$ is exactly
the same as for the crossover frequencies that were previously determined
for the vibrational DOS $g(\omega)$, since the crossover occurs exactly at
$\omega = \omega_\xi$ in Fig.~\ref{fig:wl}.  Thus, the characteristic
length $\Lambda$ can be identified with the wavelength of the modes.

Next we address the localization behavior of the vibrational excitations.
We want to know if there is an intimate relationship between the FPC and
the localization behavior, such that, for example, phonons are extended and
fractons are localized, as one might infer from the analogies with random
walks discussed above.  This analogy suggests a LDT both in $d=2$ and
$d=3$.  On the other hand there is no LDT in the somewhat related Anderson
model in $d=2$ and computer simulations for fractons and electrons (in
tight-binding approximation) on percolation clusters at $p_{\rm c}$ have
shown that the localization behavior of fractons and electrons is very
similar for corresponding frequencies and energies \cite{kantelhardt97}.
While in the quantum mechanical description the localization of the
electrons is caused by scattering and interference, the vibrational modes
are waves already in the classical description and may become localized by
scattering and interference like the electrons.

In the following we address the questions:  ({\it i\/}) Are the phonon
states in $d=2$ (for $\omega < \omega_\xi$) extended or localized states?
({\it ii\/}) Does the localization-delocalization transition (LDT) for
vibrations in $d=3$ occur at the same frequency as $\omega_\xi$?  For
answering these questions, we have applied the method of level statistics,
which proved to be a powerful tool for the electronic case recently
\cite{SEA93,HS94a,BSZK96}.  In conducting disordered electronic systems,
the energy spacing distribution $P(s)$ of consecutive eigenvalues (levels)
$E_i$ shows the universal random matrix theory result \cite{Meh91,Efe83},
which is well approximated by the appropriate Wigner surmise, $P(s) =
(\pi/2) s \exp(-\pi s^2/4)$.  Here, $s = |E_{i+1}-E_i|/\Delta$ where
$\Delta$ is the mean level spacing in the energy interval considered.  For
localized states the uncorrelated eigenvalues are described by the Poisson
distribution, $P(s) = \exp(-s)$.  In contrast to $P(s=0)=1$ for localized
states, the probability that two eigenvalues are close in energy decreases
to zero for extended states, because of the level repulsion due to the
overlap of the corresponding eigenstates.  For finite systems, the shape of
$P(s)$ is in between the two limiting cases and approaches one of them with
increasing system size.  $P(s)$ is system size independent only directly at
the transition.  Hence, the knowledge of the nearest neighbor spacing
distribution $P(s)$ for several system sizes tells whether the eigenstates
are localized or extended in the limit of infinite system size.

\begin{figure}\centering
\epsfxsize8.4cm\epsfbox{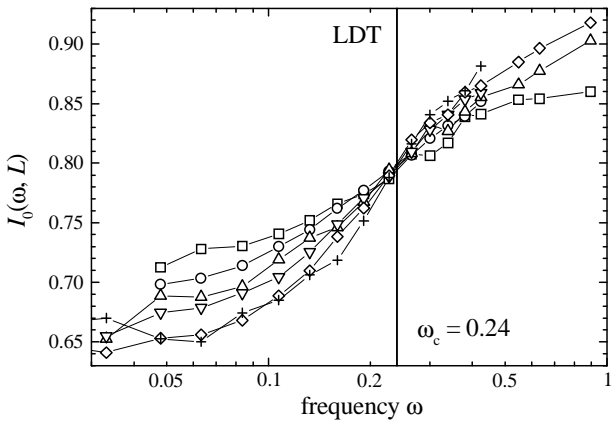}
\parbox{8.5cm}{\caption[]{\small
Second moments $I_0(\omega)$ of the level spacing distribution $P(s)$ for
vibrational modes of 3d percolation clusters with site concentration
$p=0.37$.  The symbols denote different system sizes $L$ (radius of the
clusters in topological space), $L=50$ ($\Box$), $L=62$ ({\Large$\circ$}),
$L=75$ ($\bigtriangleup$), $L=90$ ($\bigtriangledown$), $L=110$
($\Diamond$), and $L=150$ ($+$), corresponding to cubic systems with linear
sizes ranging from 30 to 107.  The correlation length was determined to be
$\xi = 3.3$.  The LDT occurs for $\omega_{\rm c} = 0.24$.}
\label{fig:I03d}}
\end{figure}

Since it is tedious to compare the $P(s)$ distributions for several system
sizes and other model parameters, we calculate the size dependence of the
quantity $I_0 = \langle s^2 \rangle /2$, which is related to the second
moment of the level spacing distribution $\langle s^2\rangle= \int_0^\infty
s^2 P(s)\,ds$.  If the eigenstates are localized, $I_0$ increases
monotonously with increasing system size approaching the Poisson limit
$I_0^{\rm loc} = 1$ for infinite system size.  If the modes are extended,
$I_0$ decreases monotonously until it reaches the random matrix theory
limit $I_0^{\rm ext} = 0.637$.  At the transition $I_0^{\rm crit}$ is scale
invariant, but its precise value depends on the boundary conditions. Other
moments of the distribution give similar results.

\begin{figure}\centering
\epsfxsize8.4cm\epsfbox{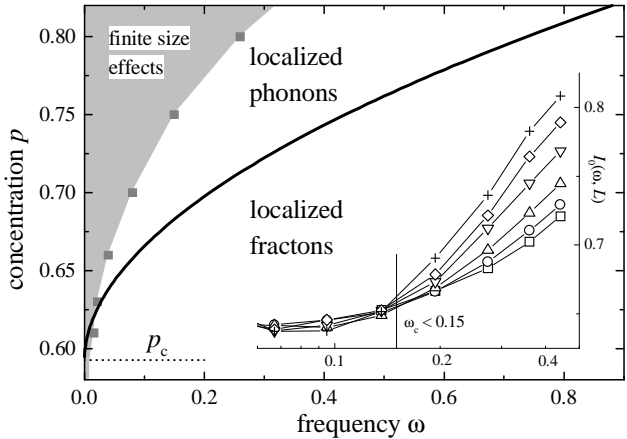}
\parbox{8.5cm}{\caption[]{\small
Phase diagram of vibrational modes of percolation clusters in $d=2$.  The
black line separates localized fractons with $\lambda(\omega) \sim
\omega^{-2/d_{\rm w}}$ from localized phonons with $\lambda(\omega) \sim
\omega^{-1}$.  In the shaded region, the method is not specific since
finite size effects occur.  The inset shows $I_0(\omega)$ for site
concentration $p=0.75$.  The symbols correspond to different system sizes,
$L=70$ ($\Box$), $L=100$ ({\Large$\circ$}), $L=150$ ($\bigtriangleup$),
$L=220$ ($\bigtriangledown$), $L=350$ ($\Diamond$), and $L=500$ ($+$),
corresponding to quadratic systems with linear sizes ranging from 95 to
680.  There is no indication of extended states.}
\label{fig:I0phas2d}}
\end{figure}

We apply the method of level statistics to vibrations of incipient infinite
percolation clusters at several concentrations ranging from $p=0.61$ to
$p=0.8$ in $d=2$, and from $p=0.32$ to $p=0.5$ in $d=3$.  Up to six
different cluster sizes are compared for each concentration.  The clusters
were generated by the Leath method and the eigenvalues were computed with a
Lanczos algorithm.  For the vibrational modes, the level spacing is
$s=|\omega^2_{i+1}-\omega^2_i|/\Delta$ since the eigenvalue $\omega^2$ in
Eq.~(\ref{eq:frak}) corresponds to the energy eigenvalue $E$ for electrons.
Representative results are shown in Fig.~\ref{fig:I03d} and in the inset of
Fig.~\ref{fig:I0phas2d}.

In Fig.~\ref{fig:I03d} the LDT can clearly be observed for vibrations of
clusters in $d=3$ at $p=0.37$.  For low frequencies, $\omega < \omega_{\rm
c}$, $I_0(\omega,L)$ decreases with increasing system size $L$, indicating
extended modes, while it increases for $\omega > \omega_{\rm c}$, as
expected for localized modes.  Directly at the critical frequency
$\omega_{\rm c} \approx 0.24$, $I_0(\omega,L)$ is independent of $L$. This
value of the LDT frequency $\omega_{\rm c}$ is well above the corresponding
FPC frequency $\omega_\xi \approx 0.053$ that we obtained from
Eq.~(\ref{eq:omegaxi}) with $\omega_0 \approx 6$.  Consequently, we
discover a novel intermediate regime of {\it extended} fracton modes for
$\omega_\xi < \omega < \omega_{\rm c}$.

\begin{figure}\centering
\epsfxsize8.4cm\epsfbox{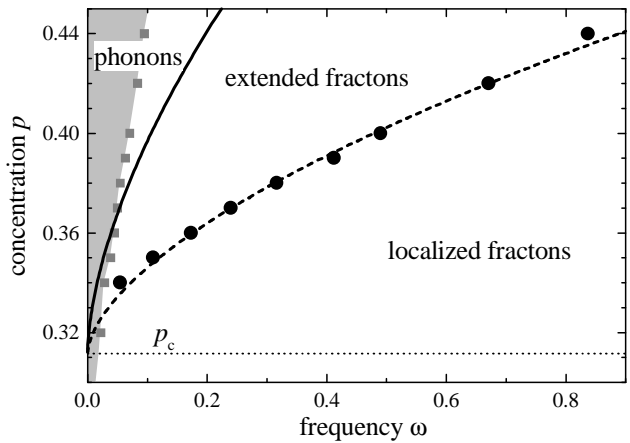}
\parbox{8.5cm}{\caption[]{\small 
Phase diagram of vibrational modes of percolation clusters in $d=3$.  The
FPC at $\omega_\xi(p)$ is marked by the continuous black line.  The black
discs indicate the LDT at $\omega_{\rm c}(p)$.  Finite size effects do only
occur in the shaded region at very low frequencies.  While the vibrational
modes are (extended) phonons for low frequencies $\omega < \omega_\xi(p)$
and localized fractons for $\omega > \omega_{\rm c}(p)$, a new intermediate
regime of extended fractons is located between the two curves.}
\label{fig:phas3d}}
\end{figure}

Using plots similar to Fig.~\ref{fig:I03d} we determined the LDT frequency
$\omega_{\rm c}(p)$ for nine concentrations $p$.  All values of
$\omega_{\rm c}(p)$ are well above the respective FPC at $\omega_\xi(p)$.
Our results for $d=3$ are summarized in Fig.~\ref{fig:phas3d} which shows
the phase diagram.  The critical frequencies $\omega_{\rm c}(p)$ can be
fitted by $\omega_{\rm c}(p) \approx 27 (p-p_{\rm c})^{d_{\rm w}\nu/2}$
(dashed line in Fig.~\ref{fig:phas3d}), suggesting that $\omega_{\rm c}(p)$
scales as $\omega_\xi(p)$, but with a different prefactor ($\omega_0
\approx 27$ instead of $\omega_0 \approx 6$).

Next we consider $d=2$.  The results of the level statistics for
vibrational modes on incipient infinite percolation clusters at $p=0.75$
are shown in the inset of Fig.~\ref{fig:I0phas2d}.  We find no indication
of a LDT.  For $\omega > 0.15$, the modes turn out to be localized since
$I_0(\omega)$ increases with increasing system size.  This limit is well
below the FPC frequency $\omega_\xi \approx 0.43$ that we obtained from
Eq.~(\ref{eq:omegaxi}) with $\omega_0 \approx 15$.  Hence, we can conclude
that the vibrational modes remain localized even for frequencies well below
the FPC frequency $\omega_\xi$.  For very small frequencies $\omega$, the
level statistics does not provide a conclusive picture due to the
restricted system sizes.  Our results for the localization behavior in
$d=2$ are summarized in the main part of Fig.~\ref{fig:I0phas2d}.

In conclusion, we have applied level statistics to investigate the
localization behavior of vibrational excitations on the infinite
percolation cluster in $d=2$ and $d=3$.  We have found that above $p_{\rm
c}$ two length scales have to be distinguished.  ({\it i\/}) The wavelength
$\lambda$ is strongly related to the density of states $g(\omega)$ and its
scaling can be derived from the analogy to diffusion.  At $\omega_\xi$,
both $\lambda(\omega)$ and $g(\omega)$ cross over from fractal to normal
behavior.  ({\it ii\/}) The localization length is supposed to diverge at
the localization-delocalization transition, which occurs in $d=3$ at a
critical frequency $\omega_{\rm c}$, that is considerably larger than
the fracton-phonon crossover frequency $\omega_\xi$. As a consequence, a
novel regime of extended fractons exists in the phase diagram of vibrations
in $d=3$. These extended states might be observable in real experiments.
In $d=2$ there is no localization-delocalization transition, so also the
phonon modes are localized.  Thus, the vibrational modes in percolation
systems show features characteristic of both classical diffusive and
electronic systems.  For a complete understanding of the problem both
analogies are useful and necessary.

We thank the Deutsche Forschungsgemeinschaft for financial support.

\end{multicols}

\vspace{-.5cm}
\begin{thebibliography}{99}

\vspace{-1cm}
\bibitem{stauffer} D. Stauffer and A. Aharony, {\it Introduction to
 Percolation Theory}, 2nd ed. (Taylor \& Francis, London, 1992).
\bibitem{bunde96} A. Bunde and S. Havlin (eds.), {\it Fractals and 
 Disordered Systems}, 2nd ed. (Springer Verlag, Heidelberg 1996).
\bibitem{alexander} S. Alexander and R. Orbach,
 J. Phys. (Paris) Lett. {\bf 43}, L-625 (1982).
\bibitem{orbach} R. Orbach, Science {\bf 231}, 814 (1986).
\bibitem{havlin} S. Havlin and D. Ben-Avraham, Adv. Phys. {\bf 36}, 695
 (1987).
\bibitem{grest} G.~S. Grest and I. Webman, 
 J. Phys. (Paris) Lett. {\bf 45}, L-1155 (1984).
\bibitem{southern} B.~W. Southern and A.~R. Douchant, 
 Phys. Rev. Lett. {\bf 55}, 966 (1985).
\bibitem{yakubo} K. Yakubo and T. Nakayama, Phys. Rev. B {\bf 40}, 517 
 (1989).
\bibitem{devries} P. de Vries, H. de Raedt, and A. Lagendijk,
 Phys. Rev. Lett. {\bf 62}, 2515 (1989); 
 H.~E. Roman, S. Russ, and A. Bunde, {\it ibid\/} {\bf 66}, 1643 (1991).
\bibitem{lambert} C.~J. Lambert and G.~D. Hughes,
 Phys. Rev. Lett. {\bf 66}, 1074 (1991).
\bibitem{bunde92} A. Bunde, H.~E. Roman, S. Russ, A. Aharony, and A.~B. 
 Harris, Phys. Rev. Lett. {\bf 69}, 3189 (1992).
\bibitem{petri} A. Petri and L. Pietronero, Phys. Rev. B {\bf 45},
 12864 (1992); A. Petri {\it et al.}, {\it ibid\/} {\bf 49}, 15067 (1994).
\bibitem{argyrakis} P. Argyrakis, S.~N. Evangelou, and M. Magoutis,
 Z. Phys. B {\bf 87}, 257 (1992); S.~N. Evangelou and P. Argyrakis,
 Phys. Rev. B {\bf 51}, 3489 (1994).
\bibitem{kantelhardt97} J.~W. Kantelhardt and A. Bunde,
 Phys. Rev. E {\bf 56}, 6693 (1997).
\bibitem{tsujimi} Y. Tsujimi, E. Courtens, J. Pelous, and R. Vacher,
 Phys. Rev. Lett. {\bf 60}, 2757 (1988); R. Vacher {\it et al.}, 
 {\it ibid} {\bf 65}, 1008 (1990).
\bibitem{montagna} M. Montagna {\it et al.}, Phys. Rev. Lett. {\bf 65}, 
 1136 (1990).
\bibitem{stoll} E. Stoll, M. Kolb, and E. Courtens,
 Phys. Rev. Lett. {\bf 68}, 2472 (1992).
\bibitem{russ} S. Russ and B. Sapoval, Phys. Rev. Lett. {\bf 73}, 1570 
 (1994).
\bibitem{nakayama} T. Nakayama, K. Yakubo, and R.~L. Orbach,
 Rev. Mod. Phys. {\bf 66}, 381 (1994).
\bibitem{grassberger} P. Grassberger, preprint cond-mat/9808095 (1998)
\bibitem{rammal} R. Rammal and G. Toulouse, 
 J. Phys. (Paris) Lett. {\bf 44}, L-13 (1983).
\bibitem{SEA93} B.~I. Shklovskii {\it et al.}, Phys. Rev. B {\bf 47}, 
 11487 (1993).
\bibitem{HS94a} E. Hofstetter and M. Schreiber, Phys. Rev. Lett. {\bf 73}, 
 3137 (1994).
\bibitem{BSZK96} M. Batsch, L. Schweitzer, I. {Kh. Zharekeshev}, 
 and B. Kramer, Phys.\ Rev.\ Lett. {\bf 77}, 1552 (1996).
\bibitem{Meh91} M.~L. Mehta, {\em Random Matrices}, 2nd ed. (Academic 
 Press, San Diego, 1991).
\bibitem{Efe83} K.~B. Efetov, Adv. Phys. {\bf 32}, 53 (1983).
\end{thebibliography}
\end{document}